\pdfoutput=1
\documentclass[10 pt]{article}

\usepackage{mathrsfs}
\usepackage{amssymb}
\usepackage{amsmath}
\usepackage{cite}
\usepackage{graphicx}
\usepackage[caption=false]{subfig}
\usepackage{color}
\usepackage{float}
\usepackage{multicol}
\usepackage{multirow}
\usepackage{soul}
\usepackage{bm}
\usepackage{enumerate}
\usepackage[margin=1.25in]{geometry}

\newtheorem{theorem}{Theorem}
\newtheorem{lemma}{Lemma}

\usepackage[table]{xcolor}
\definecolor{mygray}{gray}{.9}

\newcommand{\image}{{\rm image\ } }

\newcommand{\diag}{{\rm diag\;}}
\newcommand{\col}{{\rm col\;}}

\def\matt#1{\begin{bmatrix}#1\end{bmatrix}}

\def\eq#1{\begin{equation}#1\end{equation}}
\def\qed{ \rule{.1in}{.1in}}

\begin{document}
	
	\title{A Double-Layered Framework for Distributed Coordination in Solving Linear Equations}
	
	\author{Xuan Wang, Shaoshuai Mou, and Brian. D. O. Anderson
		\thanks{X. Wang and S. Mou are with the School of Aeronautics and Astronautics, Purdue University, West Lafayette, IN 47906 USA (e-mail: wang3156@purdue.edu, mous@purdue.edu). B. D. O. Anderson is with Hangzhou Dianzi University, Hangzhou, China,  The Australian National University and Data-61 CSIRO (formerly NICTA), Canberra, ACT 2600 Australia, {e-mail: Brian.Anderson@anu.edu.au}; his work is supported by Data-61, CSIRO, and by the Australian Research Council's Discovery Projects  DP-130103610 and DP-160104500. Corresponding Author: Shaoshuai Mou.}}


	\maketitle

	\begin{abstract}
		This paper proposes a double-layered framework (or form of network) to integrate two mechanisms, termed consensus and conservation, achieving distributed solution of a linear equation. The multi-agent framework considered in the paper is composed of clusters (which serve as a form of aggregating agent) and each cluster consists of a sub-network of agents.
		By achieving consensus and conservation through agent-agent communications in the same cluster and cluster-cluster communications, distributed algorithms are devised for agents to cooperatively achieve a solution to the overall linear equation. These algorithms outperform existing consensus-based algorithms, including but not limited to the following aspects: first, each agent does not have to know as much as a complete row or column of the overall equation; second, each agent only needs to control as few as two scalar states when the number of clusters and the number of agents are sufficiently large; third, the dimensions of agents' states in the proposed algorithms do not have to be the same (while in contrast, algorithms based on the idea of standard consensus inherently require all agents' states to be of the same dimension). Both analytical proof and simulation results are provided to validate exponential convergence of the proposed distributed algorithms in solving linear equations.
	\end{abstract}

	\thispagestyle{empty}
	\pagestyle{empty}
	
	\section{Introduction}
	Distributed control of multi-agent networks has recently received a significant amount of research attention, the goal of which is to accomplish global objectives through local coordinations \cite{FJS09DC}. \emph{Consensus}, which drives all agents in the network to reach an agreement regarding a certain quantity \cite{AJA03TAC,MAB08SIAM,RR04TAC,WR05TAC,MAB08TAC,JWH11Auto}, has served as a basis in deriving many distributed algorithms for optimization \cite{AA09TAC,JAM12TAC,TAA14TAC,XPY17TAC,QSY17TAC}, synchronization of coupled oscillators \cite{FMF13PNAS}, multi-robot formation control \cite{JR04TAC}, cooperative sensing \cite{Cortes04TRA}, and so on. Most recently consensus has motivated distributed algorithms for solving linear algebraic equations \cite{SJA15TAC,SZLDA16SCL,BSUA16NACO,XSD17TIE,GBU17TAC}, which achieve efficiency by decomposing a large system of linear equations into smaller ones that can be cooperatively solved by a network of agents. Elegant as the idea of consensus is, it has also limited application of consensus-based algorithms into situations when coordination among agents requires more than reaching consensus, especially when conservation requirements are involved. Different from consensus, \emph{conservation} is a constraint that the sum of functions of agents' states needs to be constant. Various types of conservations arise in many engineering applications including conservation of labor in ant colony \cite{G92ARE}, conservation of total energy in controlling hybrid vehicles \cite{JH08CST}, conservation of flows in traffic control \cite{MA12traffic}, conservation of linear and angular momentum in formation control \cite{ZSBC18Auto}, and so on. The idea of conservation has motivated researchers to develop algorithms for distributed resource allocations \cite{EAGJ97BCEC,CXTX17NNLS,YYQRG17WCNC}, in which all agents aim to maximize an objective function subject to a conservation constraint. Recognition of the potential of conservation in complementing consensus motives us to integrate both consensus and conservation together in one framework with the goal of combining their advantages together for achieving efficiency of distributed coordination.
	
	One natural way to achieve such integration is by layered coordination\cite{MA13PRX}, which has proved to be a powerful tool in many similar situations. For example, practical tasks involving a large number of robots can be achieved by coordination in the planning layer, executive layer and/or behavior layer. \cite{RTMDDAR02Robot,Kumar02IJRR}. Complicated optimization problems can be solved by coordination through layers, each of which iterates on its own subsets of decision variables using local information to achieve individual optimality \cite{Mung07PIEEE}. Deep learning algorithms can be established by grouping neural nodes into multiple layers to achieve different functions including feature extraction, collection, comparison, and fusion \cite{J15NN}. Compared with single-layered networks, double-layered networks (which also known multiplex networks or networks of networks \cite{QE16Nature}), provide a natural description for quantifying the interconnectivity between different categories of connections \cite{SAJ13PRL}, improve accuracy \cite{JR15NEURON}, lead to faster convergence \cite{MKKR08IFAC,IAA14NJP} and efficiency \cite{KM13TSP}. This has motivated us to employ double-layered frameworks for the integration of consensus and conservation.

	

	Similar to the architecture of mixing macro-cells and low-power nodes in mobile communication networks \cite{AJYTT11Wireless,GZB16TVT}, we consider a double-layered multi-agent network composed of clusters and each cluster consists of a network of agents. As a base station' role in communication networks, a cluster is able to communicate with other clusters and distribute information among agents within the cluster. Without losing generality we still consider communication constraints on the information flow among agents in each cluster. Such a double-layered multi-agent network allows
	the possibility that consensus is taken care of either by the layer of clusters, or the layer of agents within the same clusters, and conservation is taken care of by the other layer of the network. 
	This architecture is the basis for us to develop distributed algorithms for solving linear equations, which outperform existing consensus-based distributed linear equation solvers \cite{SJA15TAC,SZLDA16SCL,BSUA16NACO,XSD17TIE,GBU17TAC} including but not limited to the following aspects: first, each agent does not have to know as much as a complete row or column of the overall equation; second, each agent only needs to control as few as two scalar states, this being achieved  when the number of clusters and the number of agents are equal to the number of rows and columns of the overall equation, respectively; third, the dimensions of state vectors of all agents in the proposed algorithms do not have to be the same, which is in contrast to algorithms based on the idea of standard consensus. These consensus-based algorithms inherently require all agents' states to be of the same dimension. Further and differently from algorithms in \cite{MKKR08IFAC,KM13TSP,IAA14NJP}, which perform updates from lower layers to high layers in hierarchical networks for achieving consensus, the proposed algorithms do not require updates in a hierarchical order, or multi-hop communications among agents. Moreover, the layer of clusters is only responsible for communications without any burden of computations.

	\smallskip

	The rest of this paper is organized as follows.  In Section \ref{Sec_Prob} we describe the structure and information flow of a double-layer framework, and formulate the problem of solving linear equation. According to different goals in the cluster layer and the agent layer, we then present two different types of distributed algorithms for solving linear equations in Section \ref{Sec_GCLC0} and Section \ref{Sec_GCLC}, respectively. Both algorithms are distributed and converge exponentially fast. Analytical proofs and simulations are provided in these two sections. We finally conclude in Section \ref{Sec_Con} and provide proofs of lemmas in Section \ref{Sec_Appendix}.
	
	\smallskip
	\noindent{ \emph{Notation}:}
	Throughout this paper, we let ${\bf 1}_r$ denote a vector in $\mathbb{R}^r$ with all its components equal to 1; let $I_r$ denote the $r\times r$ identity matrix. Let $M'$, $\ker M$ and $\image M$ denote the conjugate transpose, the kernel and the image of a matrix $M$, respectively (Most but not all matrices and vectors will be real). Let $\col\{A_1,A_2,\cdots,A_r\}$ denote a column stack of matrices $A_i,i=1,2,...,r$, which is
	$\left[\begin{array}{cccc}
	A_1'&A_2'& \cdots& A_r'\end{array}\right]'$. Let $\diag\{A_1,A_2,\cdots,A_r\}$ denote the block diagonal matrix with $A_i$ the $i$th diagonal block entry, $i=1,2,\cdots,r$. Let $\otimes$ denote the Kronecker product.
	
	\section{Problem formulation} \label{Sec_Prob}
\begin{figure}[h!]
		\centering
		\includegraphics[width=7 cm]{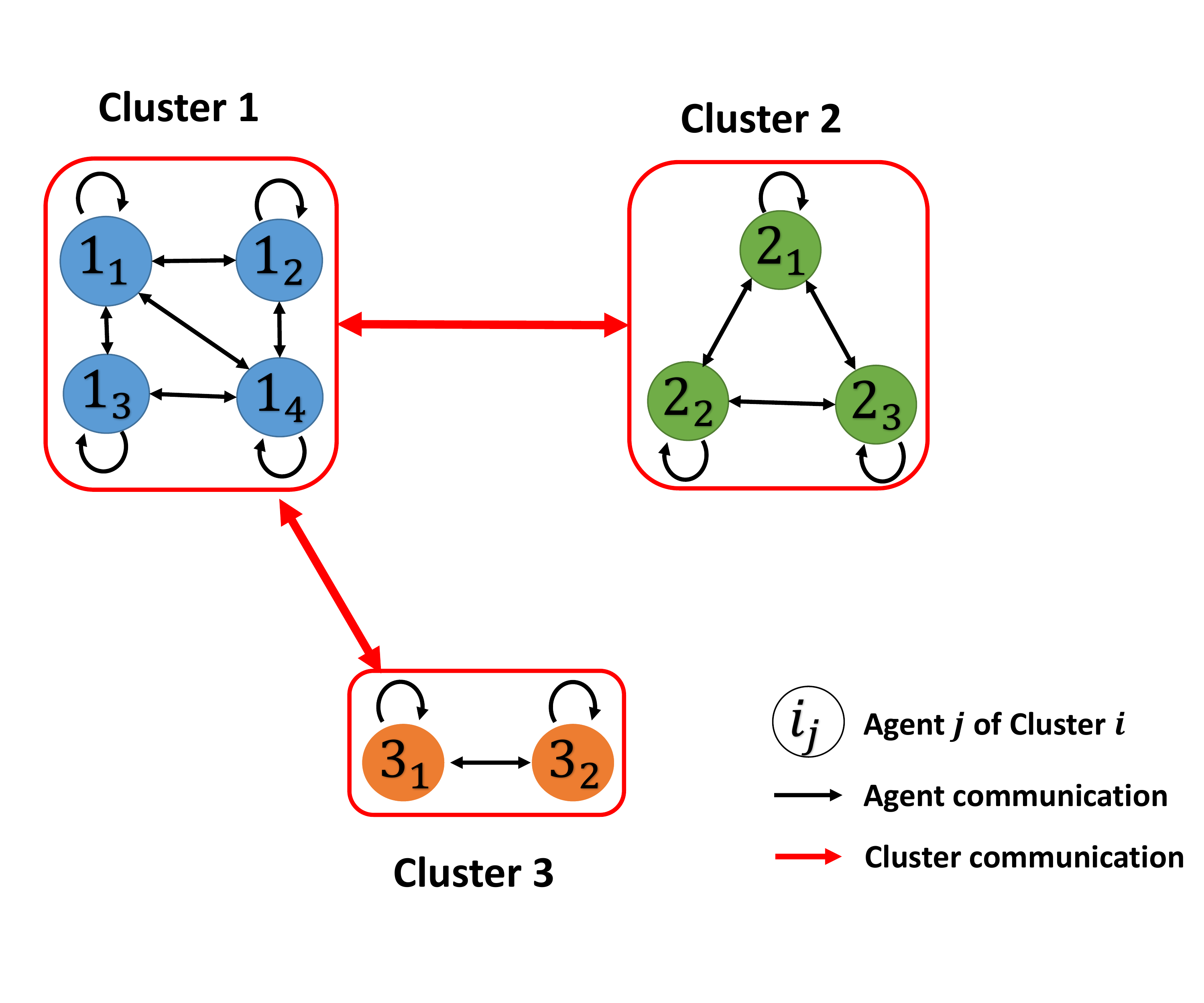}
		\caption{An Example of a Double-layered Multi-Agent Network }
		\label{Fig1}
	\end{figure}

	Consider a double-layer multi-agent network consisting of a number of $c$ clusters, the $i$-th of which is composed of a number of $c_i$ agents. Suppose each cluster $i$, $i=1,\cdots,c$, is able to receive information from certain  clusters which are called $i$'s \textit{cluster-neighbors} denoted by $\mathcal{N}_i$, specially we assume $i\in\mathcal{N}_i$. The neighbor relations of clusters can be characterized by a $c$-node graph $\mathbb{G}$, in which there is an arc from $\bar{i}$ to $i$ if and only if $\bar{i}\in\mathcal{N}_i$. Within each cluster $i$, $i=1,\cdots,c$, each agent $i_j$, $j=1,\cdots,c_i$, is able to receive information from certain  agents, which are called agent $i_j$'s \textit{agent-neighbors} denoted by $\mathcal{N}_{ij}$, specially $i_{{j}}\in\mathcal{N}_{ij}$. The neighbor relations in cluster $i$ can be characterized by a $c_i$-node graph $\mathbb{G}_i$, in which there is an arc from $i_{\bar{j}}$ to $i_j$ if and only if  $i_{\bar{j}}\in\mathcal{N}_{ij}$. Suppose all $\mathbb{G}_i$, $i=1,2,...,c$ and $\mathbb{G}$ are connected and bidirectional. \footnote{Here we use bidirectional instead of undirected to emphasize the two-way information flows.} One example of such a double-layer multi-agent network is shown in Fig. \ref{Fig1}.
	
	Consider an overall linear equation $$A\bm{x}=b$$
	where $A\in\mathbb{R}^{m \times n}$ and $b\in\mathbb{R}^{m}$. Suppose $Ax=b$ has at least one solution. Suppose each agent $i_j$ knows part of the overall linear equation, which might not be as much as a complete  row or column of $A$, say. Let each agent $i_j$ control a state vector $x_{ij}(t)$, while each cluster here does not control any state and only plays the role of communications. The \textbf{problem} of interest is to develop a distributed update for each agent such that all $x_{ij}(t)$ converge to constant vectors $x_{ij}^*$, $j=1,2,...,c_i$ and $i=1,2,...,c$,  which jointly form a solution to $Ax=b$.
	
	\section{Global-Consensus and Local-Conservation} \label{Sec_GCLC0}
\begin{figure}[H]
		\centering
		\includegraphics[width=8.5 cm]{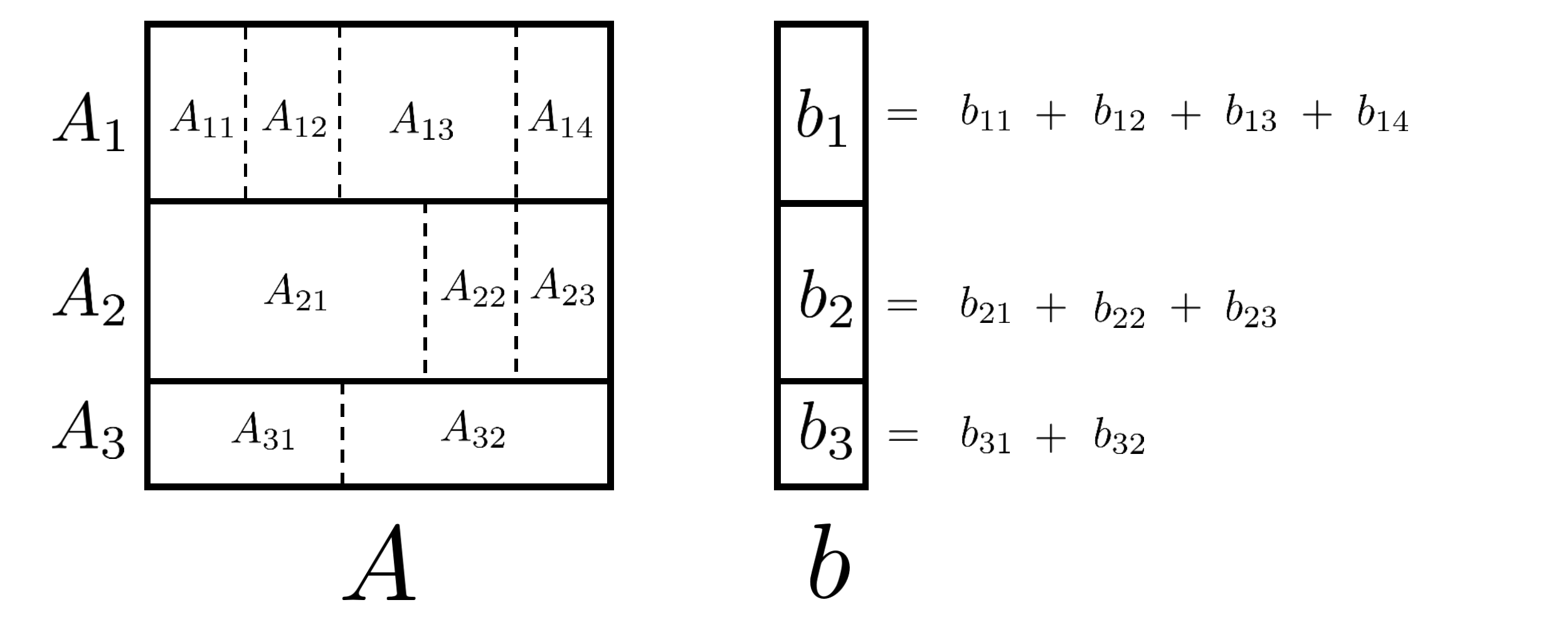}
		\caption{An example of the relation between agents' locally available information and the overall equation}
		\label{S1Demo}
	\end{figure}

	Suppose each agent $i_j$ in a cluster $i$ knows ${A}_{ij}\in\mathbb{R}^{m_i\times n_{ij}}$ and ${b}_{ij}\in\mathbb{R}^{m_i}$ such that the collection of them
	\eq{\label{eq_partition12}\begin{bmatrix}
			{A}_{i1} & {A}_{i2} & \cdots & {A}_{i{c_i}}
		\end{bmatrix}=A_i\in\mathbb{R}^{m_i\times n},\quad \sum_{j=1}^{c_i}{b}_{ij}=b_i\in\mathbb{R}^{m_i}} are a block row of the overall equation, where \begin{align}
	A=\begin{bmatrix}
	{A}_1\\
	{A}_2\\
	\vdots\\
	{A}_c
	\end{bmatrix},
	\quad
	b=\begin{bmatrix}
	{b}_1 \\ {b}_2 \\ \vdots \\ {b}_c
	\end{bmatrix}. \label{eq_partition11}
	\end{align}
	Here, one has
	\begin{align}\label{n_im_i1}
	\sum_{j=1}^{c_i}n_{ij}=n,\  i=1,2,...,c, \quad \sum_{i=1}^cm_i=m.
	\end{align}
	Note that $n_{ij}=1$ and $m_i=1$ are permitted, but are of course not required. Consistent with the set-up of Fig. \ref{Fig1}, an example of how each agent's locally available information $A_{ij}, b_{ij}$ is related to the overall equation $Ax=b$ is shown in Fig. \ref{S1Demo}.
	\noindent
	
	Suppose each agent $i_j$ controls a state vector ${x}_{ij}(t)\in\mathbb{R}^{n_{ij}}$. In this section, we aim to devise a distributed update for each agent $i_j$'s state $x_{ij}(t)$ to converge exponentially fast to a constant vector $x_{ij}^*$ such that:
	\begin{itemize}
		\item All $x_{ij}^*$, $j=1,2,...,c_i$ within each cluster $i$, $i=1,2,...,c$, satisfy the following
		\begin{align}\label{eq_conserv1}
		\text{\bf Local Conservation: }\quad &\sum_{j=1}^{c_i} ({A}_{ij}{x}^*_{ij}-b_{ij})=0.
		\end{align}
		\item All $\bm{x}^*_i=\col\{{x}^*_{i1}, \cdots, {x}^*_{ic_i}\}$, $i=1,2,...,c$, among all clusters in the network reach a consensus $\bm x^*$, that is,
		\eq{\text{\bf Global Consensus: }\quad  {\bm{x}}^*_1={\bm{x}}^*_2=\cdots={\bm{x}}^*_c=\bm{x}^* \label{Consensus} }
	\end{itemize}
	From (\ref{eq_partition12}) and (\ref{eq_conserv1}) one has ${A}_i{\bm{x}}^*_i={b}_i$. This and the global consensus in (\ref{Consensus}) imply $A\bm x^*=b.$  All $x^*_{ij}$ satisfying the local conservation (\ref{eq_conserv1}) and the global consensus (\ref{Consensus}) are said to form
	a\emph{ consensus-conservation solution} $\bm x^*$ to $Ax=b$.
	
	\subsection{The Update}
	Let $\bm{x}_i(t)\in \mathbb{R}^n$ denote a column collection of agent states in cluster $i$, $ i=1,2,...,c$, that is, \eq{\label{compxi1}\bm{x}_i(t)=\col\{{x}_{i1}(t), \cdots, {x}_{ic_i}(t)\}.}  To achieve a consensus-conservation solution $\bm {x}^*$, it is sufficient to achieve $A_i\bm{x}_i(t)=b_i$ while all $\bm{x}_i(t)$, $i=1,2,...,c$, reach a consensus. In order to decompose the global consensus of $\bm{x}_i(t)$ into relations involving agents' states, we let $E_{ij}\in \mathbb{R}^{n_{ij}\times n}$ denote a matrix consisting of rows from $I_n$ such that $\col \{E_{ij}, j=1,2,...,c_i\}=I_n$ and $$A_{ij}=A_i E_{ij}'.$$ Then one  has
	$$x_{ij}(t)=E_{ij}\bm{x}_i(t).$$ It follows that all $\bm{x}_i(t)$ reaching a consensus is equivalent to requiring that $\forall i=1,\cdots,c_i$ and $\forall k\in \mathcal{N}_i$,
	\begin{align}\label{eq_consensus}
	{x}_{ij}(t)\to E_{ij}\bm{x}_{k}(t).
	\end{align} To achieve the local conservation \eqref{eq_conserv1}, one also introduces an additional \emph{coordination state} $z_{ij}(t)\in \mathbb{R}^{m_i}$ associated with and stored by each agent $i_j$.
	
	Suppose each cluster $i$ is able to access $\bm{x}_k(t)$, $k\in \mathcal{N}_i$ through cluster-cluster communications and then distribute $E_{ij}\bm{x}_k(t)$ to each agent $i_j$ within the cluster $i$. Within each cluster $i$, each agent $i_j$ is able to access its neighbors' coordination state $z_{ik}, i_k\in \mathcal{N}_{ij},$ through agent-agent communication in cluster $i$. Then we propose the following update for each agent $i_j$, $i=1,2,...,c$ and $j=1,2,...,c_i$, at time $t$:
	\begin{align}	\label{LEFlowx}
	\dot{x}_{ij}=&-A_{ij}' \left(A_{ij}x_{ij}-b_{ij}-\sum\limits_{i_k \in {\mathcal{N}_{ij}}} \left (z_{ij}-z_{ik}\right)\right)\nonumber\\
	&-\sum_{k\in \mathcal{N}_i}\left({x}_{ij}-E_{ij}\bm x_k\right)\\
	\dot{z}_{ij}=&A_{ij}x_{ij}-b_{ij}-\sum\limits_{i_k \in {\mathcal{N}_{ij}}} \left (z_{ij}-z_{ik}\right) \label{LEFlowz}
	\end{align}
	where the first line of update \eqref{LEFlowx} and \eqref{LEFlowz} aim to achieve the local conservation \eqref{eq_conserv1} while the second line of update \eqref{LEFlowx} aims to achieve the global-consensus (\ref{Consensus}). One natural generalization to the proposed updates \eqref{LEFlowx}-\eqref{LEFlowz} is achievable by assigning different weights to controls for the local conservation and the global consensus, respectively, with the aim of achieving faster convergence. Optimal choice of such weights might however require more than locally available information, which will not be discussed in this paper.
	
	Note immediately that the proposed updates \eqref{LEFlowx}-\eqref{LEFlowz} are \textbf{distributed} in the sense that implementation of them only require communications of states $\bm{x}_k(t)$ among cluster-neighbors and communications of coordination states $z_{ik}(t)$ among neighbor agents within the same cluster. Moreover, compared with existing consensus-based distributed linear equation solvers \cite{SJA15TAC,SZLDA16SCL,BSUA16NACO,XSD17TIE,GBU17TAC}, distributed updates \eqref{LEFlowx}-\eqref{LEFlowz}
	\begin{itemize}
		\item require \textbf{much less knowledge} of the overall equation and control states of \textbf{much smaller dimension}. For a given $A\in \mathbb{R}^{m\times n}$ of fixed size, each agent $i_j$ knows $A_{ij}\in \mathbb{R}^{m_i\times n_{ij}}$ and $b_{ij}\in \mathbb{R}^{m_i}$, and controls states $x_{ij}(t)\in \mathbb{R}^{n_{ij}}$, and $z_{ij}(t)\in \mathbb{R}^{m_i}$. Sizes of all these locally available matrices and state vectors might change with respect to the number of clusters and the number of agents in each cluster. To see why this is so, we note from \eqref{n_im_i1} and partitions in Fig. \ref{S1Demo} that increasing $c$ and $c_i$ leads to the decreases of $m_i$ and $n_{ij}$, respectively. Specially, when the number of clusters is $m$ and the number of agents within each cluster is $n$, that is, $c=m$ and $c_i=n$, each agent only needs to know \textbf{two scalar entries} $A_{ij}\in \mathbb{R}$, $b_{ij}\in \mathbb{R}$ and  updates \textbf{two scalar states}, namely $x_{ij}(t)\in \mathbb{R}, \ z_{ij}(t)\in \mathbb{R}$.
		\item allow all agents' state vectors to be of \textbf{different dimensions} while in contrast consensus-based distributed linear equation solvers require all agents to control states of the same size. Thus the proposed updates might be applied in networks of heterogeneous agents with different capability of storage.
	\end{itemize}


	\subsection{Main result}
	Before proceeding, we first derive a compact form of \eqref{LEFlowx}-\eqref{LEFlowz}. Towards this end, we let $\bm{z}_i(t)\in \mathbb{R}^{c_im_i}$ denote the column collection of all agents' coordination states in cluster $i$, that is,
	\eq{\bm{z}_i(t)=\col\{{z}_{i1}(t), \cdots, {z}_{ic_i}(t)\}, \quad i=1,2,...,c.} Let \eq{\label{eq_diag}\bar{A}_{i}=\diag\{{A}_{i1}, \cdots,  {A}_{i{c_i}}\}, \quad \bar{b}_{i}=\col\{{b}_{i1},\cdots, {b}_{i{c_i}}\}} and \eq{\label{eq_bl}\bar{L}_{\mathbb{G}_i}={L}_{\mathbb{G}_i}\otimes I_{m_i},\quad i=1,2,...,c}
	with ${L}_{\mathbb{G}_i}$ the Laplacian matrix of the $c_i$-node connected and bidirectional graph $\mathbb{G}_i$. Recalling $\col \{E_{ij}, j=1,2,...,c_i\}=I_n$ and \eqref{compxi1}, one can write equations \eqref{LEFlowx} and \eqref{LEFlowz} as:
	\begin{align}	\label{CCompactFlowx}
	\dot{\bm{x}}_{i}=&-\bar{A}_{i}' \left(\bar{A}_{i}\bm{x}_{i}-\bar{b}_{i}- \bar{L}_{\mathbb{G}_i}\bm{z}_i\right)-\sum_{k\in \mathcal{N}_i}\left(\bm x_i- \bm x_k\right)\\
	\dot{\bm{z}}_{i}=&\bar{A}_{i}\bm{x}_{i}-\bar{b}_{i}- \bar{L}_{\mathbb{G}_i}\bm{z}_i \label{CCompactFlowz}
	\end{align} for $i=1,2,...,c$. Each equation pair in the above describes what is occurring at a particular cluster. Now
	let $\bm{x}=\col\{\bm{x}_1, \cdots, \bm{x}_c\}$, $\bm{z}=\col\{\bm{z}_1, \cdots, \bm{z}_c\}$, \begin{align}
	\hat{A}=\diag\{\bar{A}_{i}, \cdots,  \bar{A}_{c}\}, \quad \hat{b}=\col\{\bar{b}_{i}, \cdots,  \bar{b}_{c}\},\label{eq_Ahat1}\\
	\hat{L}=\diag\{\bar{L}_{\mathbb{G}_1},\cdots,\bar{L}_{\mathbb{G}_c}\}, \quad \hat{L}_{\mathbb{G}}= {L}_{\mathbb{G}}\otimes I_{n}\label{eq_Ahat2} 
	\end{align} with  ${L}_{\mathbb{G}}$ Laplacian matrix of the $c$-node connected graph $\mathbb{G}$. Equations (\ref{CCompactFlowx})-(\ref{CCompactFlowz}) can be further rewritten in the following compact form, which describes the behavior of the whole network:
	\begin{align}	\label{WCompactFlowx}
	\dot{\bm{x}}=&-\hat{A}' \left(\hat{A}\bm{x}-\hat{b}- \hat{L}\bm{z}\right)-\hat{L}_{\mathbb{G}}\bm{x}\\
	\dot{\bm{z}}=&~\hat{A}\bm{x}-\hat{b}- \hat{L}\bm{z} \label{WCompactFlowz}
	\end{align} which is
	\begin{align}	\label{MCompact}
	\begin{bmatrix}
	\dot{\bm{x}}\\
	\dot{\bm{z}}
	\end{bmatrix}=Q\begin{bmatrix}
	{\bm{x}}\\
	{\bm{z}}
	\end{bmatrix}+\begin{bmatrix}
	\hat{A}'\hat{b}\\
	-\hat{b}
	\end{bmatrix}
	\end{align}
	with
	\begin{align} \label{eq_Q1}
	Q=\begin{bmatrix}
	-\hat{A}'\hat{A}-\hat{L}_{\mathbb{G}} & \hat{A}' \hat{L}\\ \\
	\hat{A} & -\hat{L}
	\end{bmatrix}.
	\end{align}
	To analyze the convergence of (\ref{MCompact}) we need the following lemma to characterize eigenvalues of $Q$.
	%
	%
	%
	%

	\begin{lemma}\label{Lemma_eig}
		Let
		$$M=\begin{bmatrix}
		-M_1'M_1-M_2 & M_1' M_3\\ \\
		M_1 & -M_3
		\end{bmatrix}$$
		where the $M_i$ are real, $i=1,2,3$, and $M_2$ and $M_3$ are positive semi-definite. Then all eigenvalues of $M$ are real negative or 0. Moreover, if $0$ is an eigenvalue of $M$, it must be non-defective\footnote{ An eigenvalue is non-defective if any only if its algebraic multiplicity equals its geometric multiplicity. In other words, the Jordan block corresponding to a non-defective eigenvalue is diagonal.}.
	\end{lemma}
	\smallskip
	
	The proof of Lemma \ref{Lemma_eig} will be given in the Appendix. By this lemma and by establishing the convergence of the linear time-invariant system (\ref{MCompact}) to a constant steady state, one has the following main result.
	
	\begin{theorem}\label{T1}
		Suppose $Ax=b$ has at least one solution, and the graphs $\mathbb{G}_i$, $i=1,2,...,c$, $\mathbb{G}$ are  connected and bidirectional. Then under the distributed updates \eqref{LEFlowx}-\eqref{LEFlowz}, all $x_{ij}(t)$ with $i=1,2,\cdots,c$ and $j=1,2,...,c_i$ converge exponentially fast to constant vectors $x_{ij}^*$ satisfying (\ref{eq_conserv1})-(\ref{Consensus}), which form a consensus-conservation solution $\bm x^*$ to $Ax=b$.
	\end{theorem}
	\smallskip

	\noindent\textbf{Proof of Theorem \ref{T1}:} We first prove that there exists a constant vector $\col\{\hat{\bm{x}},\hat{\bm{z}}\}$ which is an equilibrium of \eqref{MCompact}. Recall there exists a constant vector $y\in \mathbb{R}^n$ such that $Ay=b$. From the definitions of $A_{ij}, b_{ij}$ in (\ref{eq_partition12})-(\ref{eq_partition11}) and $E_{ij}$, one has $$\sum_{j=1}^{c_i} ({A}_{ij}E_{ij}y-b_{ij})=0, \quad i=1,2,...,c,$$
	This equation and definitions of $\bar{A}_i, \bar{b}_i$ in \eqref{eq_diag} lead to \eq{\label{xespace}\left({\bf 1}'_{c_i}\otimes I_{m_i}\right)\left(\bar{A}_i y-\bar{b}_i\right)=0,\quad i=1,2,...,c.} Note that $\bar{L}_{\mathbb{G}_i}=L_{\mathbb{G}_i}\otimes I_{m_i}$, where $L_{\mathbb{G}_i}$ is the Laplacian matrix of a $c_i$-node connected and bidirectional graph $\mathbb{G}_i$. Then \eq{\label{eq_imageL1}\image \bar{L}_{\mathbb{G}_i}=\ker \left({\bf 1}'_{c_i}\otimes I_{m_i}\right),\quad i=1,2,...,c} From \eqref{xespace} and (\ref{eq_imageL1}), one has
	\begin{align}\label{xespace2}
	\left(\bar{A}_i y-\bar{b}_i\right)\in \image \bar{L}_{\mathbb{G}_i}, \quad i=1,2,...,c.
	\end{align} Then there exists a constant vector $\hat{z}_i\in \mathbb{R}^{c_im_i}$ such that
	\begin{align} \label{xezeeql}
	\bar{A}_iy-\bar{b}_i-\bar{L}_{\mathbb{G}_i}\hat{z}_i=0, \quad i=1,2,...,c.
	\end{align} Let $\hat{\bm{x}}={\bm 1}_c\otimes y$. Note that $\hat{L}_{\mathbb{G}}= {L}_{\mathbb{G}}\otimes I_{n}$ with ${L}_{\mathbb{G}}$ the Laplacian matrix of the $c$-node connected and bidirectional graph $\mathbb{G}$. Then \eq{\label{eq_Lbx}\hat{L}_{\mathbb{G}}\hat{\bm{x}}=0} Let $\hat{\bm z}=\{\hat{z}_1,\hat{z}_2,...,\hat{z}_c\}$. From (\ref{xezeeql}) and definitions of $\hat{A},\hat{b}$ in (\ref{eq_Ahat1}), one has \eq{\hat{A}\hat{\bm{x}}-\hat{b}- \hat{L}\hat{\bm{z}}=0,}
	This equation and (\ref{eq_Lbx}) imply that $\col\{\hat{\bm x},\hat{\bm z}\}$ is an equilibrium of \eqref{MCompact}.
	\smallskip

	Second, we analyze the convergence of the error
	\begin{align}
	\bm{e}(t)=\begin{bmatrix}
	\bm{x}(t)\\
	\bm{z}(t)
	\end{bmatrix}-\begin{bmatrix}
	\hat{\bm x}\\
	\hat{\bm z}
	\end{bmatrix}.
	\end{align}
	From \eqref{MCompact} and the fact that $\col\{\hat{\bm x},\hat{\bm z}\}$ is an equilibrium of \eqref{MCompact}, one has
	\begin{align}\label{eupdate}
	\bm{\dot e}=Q\bm{e}
	\end{align}
	From Lemma \ref{Lemma_eig}, the structure of $Q$ in (\ref{eq_Q1}) and the fact that the Laplacian matrices $\hat{L}$ and $\hat{L}_{\mathbb{G}}$ are symmetric and positive semi-definite, one concludes that all eigenvalues of $Q$ are real negative or 0. Moreover, if $0$ is an eigenvalue of $Q$, it must be non-defective. Thus there exists a constant vector $q\in \ker Q$ such that $\bm{e}(t)$ of the linear time-invariant system (\ref{eupdate}) converges to $q$ exponentially fast\cite{HAMLinearSys}. Thus  $\col \{\hat{\bm x}(t),\hat{\bm z}(t)\}$ converges exponentially fast to a constant vector $\col\{\hat{\bm x}^*, \hat{\bm z}^*\}$, where
	\eq{\label{eq_xzs1}\begin{bmatrix}
			\hat{\bm x}^*\\
			\hat{\bm z}^*
		\end{bmatrix}=\begin{bmatrix}
		\hat{\bm x}\\
		\hat{\bm z}
	\end{bmatrix}
	+q,\quad q\in \ker Q.} Partition the constant vector $\hat{\bm x}^*$ such that \eq{\label{eq_xx1}\hat{\bm{x}}^*=\col\{{\bm x}^*_1, \cdots, {\bm x}^*_{c}\}} where $\bm{x}_i^*\in \mathbb{R}^n$ is further partitioned as \eq{\bm{x}_i^*=\col \{x_{i1}^*,x_{i2}^*,...,x_{ic_i}^*\} \label{eq_xxx}}
with $x_{ij}^*\in \mathbb{R}^{n_{ij}}$. Evidently, we have that $x_{ij}(t)$ converges to $x_{ij}^*$ exponentially fast. In the following one only needs to show that all these $x_{ij}^*$ satisfy the local conservation in (\ref{eq_conserv1}) and the global consensus in (\ref{Consensus}).

From (\ref{eq_xzs1}) and the property that $\col\{\hat{\bm x},\hat{\bm z}\}$ is an equilibrium of \eqref{MCompact}, one concludes that $\col\{\hat{\bm x}^*,\hat{\bm z}^*\}$ is also an equilibrium of \eqref{MCompact}. It follows that
\begin{align}	\label{WCompactFlowx1}
0=&-\hat{A}' \left(\hat{A}\hat{\bm{x}}^*-\hat{b}- \hat{L}\hat{\bm z}^*\right)-\hat{L}_{\mathbb{G}}\hat{\bm x}^*\\
0=&~\hat{A}\hat{\bm x}^*-\hat{b}- \hat{L}\hat{\bm z}^* \label{WCompactFlowz1}
\end{align} Partition $\hat{\bm z}^*=\col \{\bm z_1^*,\bm z_2^*,...,\bm z_c^*\} $ with $\bm{z}_i^*\in \mathbb{R}^{c_im_i}$.  From (\ref{WCompactFlowz1}) and the definitions of $\hat{A}, \hat{b}, \hat{L}$ in (\ref{eq_Ahat1})-(\ref{eq_Ahat2}), one has \eq{\label{eq_equizi}\bar{A}_i\bm x_i^*-\bar{b}_i-\bar{L}_{\mathbb{G}_i}\bm z_i^*=0, \quad i=1,2,...,c.} From the definitions of $\bar{A}_i, \bar{b}_i, \bar{L}_{\mathbb{G}_i}$ in (\ref{eq_diag})-(\ref{eq_bl}), one can rewrite (\ref{eq_equizi}) as \eq{\label{eq_lemma1}\matt{A_{i1} x_{i1}^*\\A_{i2} x_{i2}^*\\ \vdots \\ A_{ic_i}  x_{ic_i}^*}-\matt{b_{i1}\\b_{i2}\\\vdots \\b_{ic_i}}- (L_{\mathbb{G}_i}\otimes I_{m_i})\bm z_i^*=0.}
Premultiplying by ${\bf 1}'_{c_i}\otimes I_{m_i}$ on both sides of  (\ref{eq_lemma1}), one has $$\sum_{j=1}^{c_i}(A_{ij}x_{ij}^*-b_{ij})-[({\bf 1}'_{c_i}L_{\mathbb{G}_i})\otimes I_{m_i}]\bm z_i^*=0. $$
Since $L_{\mathbb{G}_i}$ is the Laplacian of $c_i$-node connected bidirectional graph $\mathbb{G}_i$, one has ${\bf 1}'_{c_i}L_{\mathbb{G}_i}=0$. Thus
\eq{\sum_{j=1}^{c_i}(A_{ij}x_{ij}^*-b_{ij})=0. \label{eq_conserv10}}
In addition, from (\ref{WCompactFlowx1})-(\ref{WCompactFlowz1}), one has $$\hat{L}_{\mathbb{G}}\hat{\bm x}^*=0$$ where $\hat{L}_{\mathbb{G}}=L_{\mathbb{G}}\otimes I_m$ with $L_{\mathbb{G}}$ the Laplacian matrix of the $c$-node connected and bidirectional graph $\mathbb{G}$. Thus there exists a constant vector $\bm x^*$ such that \eq{\label{eq_hatbx}\hat{\bm x}^*={\bm 1}_c\otimes \bm x^*}
Together with (\ref{eq_xx1}), this implies
\eq{{\bm{x}}^*_1={\bm{x}}^*_2=\cdots={\bm{x}}^*_c=\bm{x}^* \label{eq_consensus0}}
with $\bm{x}_i^*$ a collection of $x_{ij}^*$ as defined in (\ref{eq_xxx}).
From (\ref{eq_conserv10}) and (\ref{eq_consensus0}) one concludes that all $x_{ij}^*$ satisfy the local conservation in (\ref{eq_conserv1}) and the global consensus in (\ref{Consensus}).

Therefore, the $x_{ij}(t)$ converge exponentially fast to constant vectors $x_{ij}^*$ which form a consensus-conservation solution $\bm{x}^*$ to $Ax=b$. This completes  the proof. \qed

\subsection{Simulation}
We utilize the double-layer network as in Fig. \ref{Fig1} to solve the linear equation $A\bm{x}=b$, which is partitioned according to the structure as in Fig. \ref{S1Demo} with details as follows:

\begin{figure}[H]
	\centering
	\includegraphics[width=8cm]{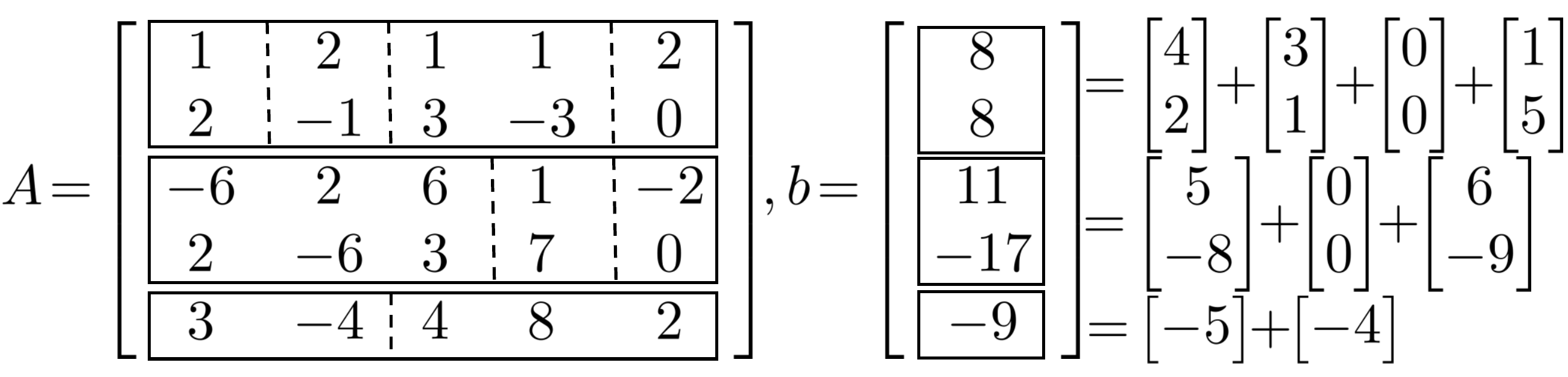}
\end{figure}

\begin{figure}[H]
	\centering
	\includegraphics[width=8.5 cm]{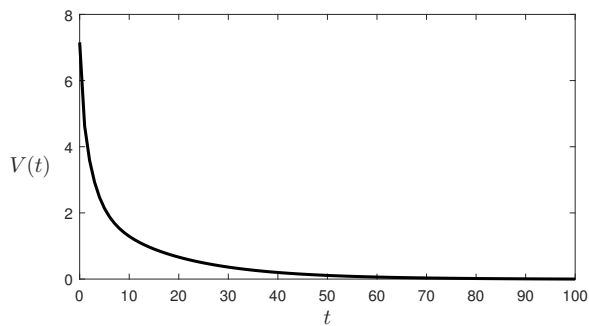}
	\caption{Evolution of $V(t)$ under the proposed updates \eqref{LEFlowx}-\eqref{LEFlowz}}
	\label{Fig3}
\end{figure}

Suppose each agent $i_j$ knows $A_{ij}$ and $b_{ij}$, and employ the updates \eqref{LEFlowx} and \eqref{LEFlowz} with arbitrary initializations. Let
$$V(t)=\dfrac{1}{2}\sum_{i=1}^c\left\|\begin{bmatrix}
x_{i1}(t)\\\vdots\\x_{ic_i}(t)
\end{bmatrix}-\bm{x}^*\right\|^2_2$$ where $\bm{x}^*=\begin{bmatrix} 1.27&3.23&2.02&-0.88&-0.43 \end{bmatrix}'$ is a solution to $Ax=b$. Thus $V(t)$ measures the closeness of all agent states to forming a consensus-conservation solution. Simulations shown in Fig. \ref{Fig3} suggest that $V(t)$ converges exponentially fast to 0, which indicates all $x_{ij}(t)$ converge exponentially fast to constant vectors that form a consensus-conservation solution $\bm x^*$. This is in accord with Theorem \ref{T1}.

\section{Global-Conservation and Local-Consensus} \label{Sec_GCLC}
In the previous section, agents in the same cluster jointly know a  block row of the overall matrix $A$ as indicated in Fig. \ref{S1Demo}. In this section, we consider a different situation in which agents in the same cluster jointly know a block column  of $A$, for which different coordination will be required in the cluster-layer and the agent-layer as will be shown later.

\begin{figure}[h]
	\vspace{-0.3cm}
	\centering
	\includegraphics[width=8.5 cm]{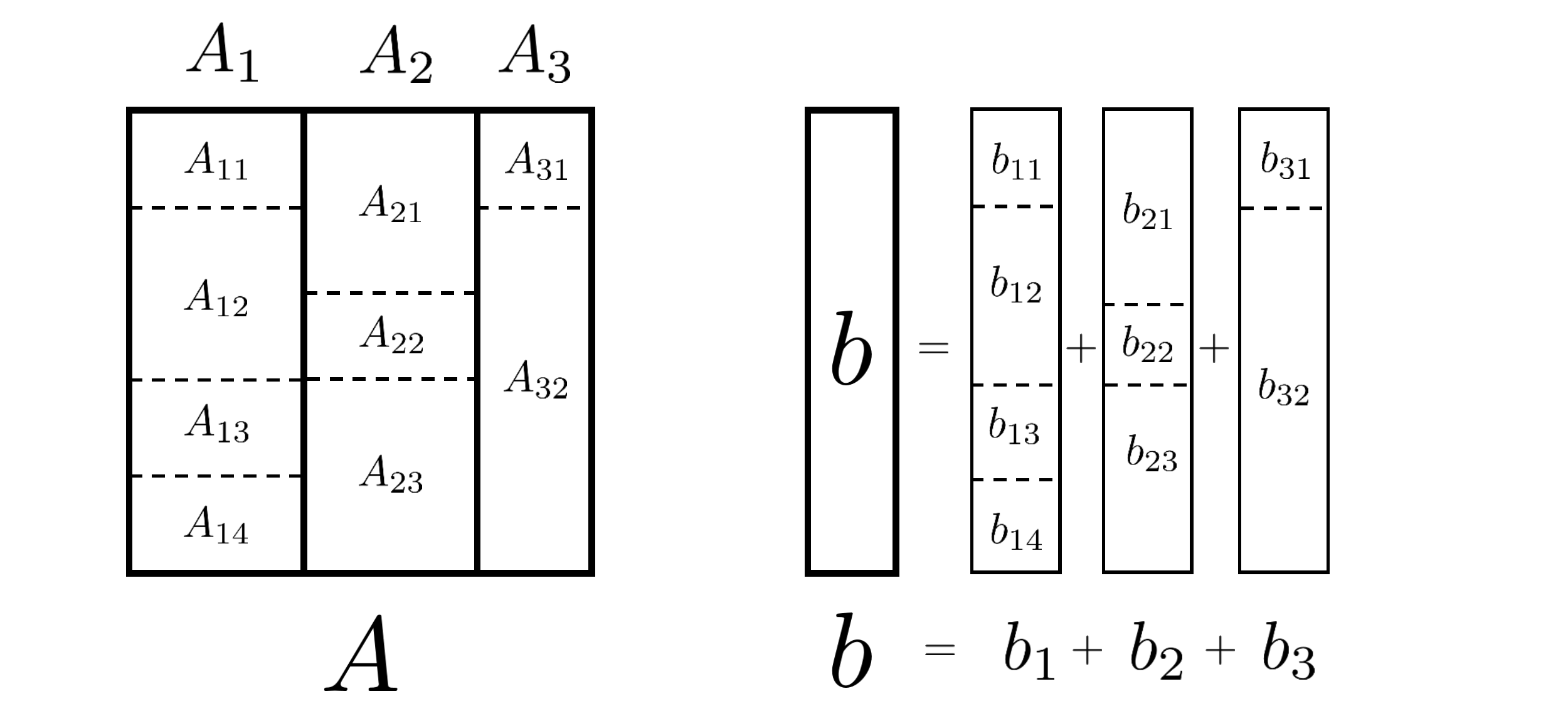}
	\caption{An example of the relation between agents' locally available information and the overall equation.}
	\label{S2Demo}
\end{figure}

Suppose each agent $i_j$ in cluster $i$ knows $A_{ij}\in\mathbb{R}^{m_{ij}\times n_i}, b_{ij}\in\mathbb{R}^{m_{ij}}$ such that the collection of them
\eq{\label{eq_partition22}\begin{bmatrix}
		{A}_{i1}\\
		{A}_{i2}\\
		\vdots\\
		{A}_{ic_i}
	\end{bmatrix}=A_i\in\mathbb{R}^{m\times n_{i}}
	\quad\quad
	\begin{bmatrix}
		{b}_{i1}\\
		{b}_{i2}\\
		\vdots\\
		{b}_{ic_i}
	\end{bmatrix}=b_i\in\mathbb{R}^{m}} are parts of the overall linear equation $Ax=b$, where \begin{align}
A=\begin{bmatrix}
{A}_{1} & {A}_{2} & \cdots & A_{c}
\end{bmatrix},\quad b=\sum_{i=1}^{c}{b}_{i}. \label{eq_partition21}
\end{align}
Then one has
\begin{align}\label{n_im_i2}
\sum_{j=1}^{c_i}m_{ij}=m,\ i=1,2,...,c,\quad \sum_{i=1}^c n_i=n.
\end{align}
An example of the relation between agents' locally available information $A_{ij}, b_{ij}$ and the overall equation $Ax=b$ is shown in Fig. \ref{S2Demo}. Note that the symbols $A_{ij}$ and $b_i$ here are used differently from their use in previous sections but the notation is convenient.
\noindent

Suppose each agent each agent $i_j$ controls a state vector ${x}_{ij}(t)\in\mathbb{R}^{n_{i}}$. In this section, we aim to devise a distributed update for each agent $i_j$'s state $x_{ij}(t)$ to converge exponentially fast to a constant vector $x_{ij}^*$, $i=1,2,...,c$ and $j=1,2,...,c_i$, such that
\begin{itemize}
	\item All $x^*_{ij}$, $j=1,2,...,c_i$, within each cluster $i$ reach a consensus $ x^*_{i}$, that is,
	\eq{\text{\bf Local Consensus: }\quad  x^*_{i1}=x^*_{i2}=\cdots=x^*_{ic_i}=x^*_i \label{Consensus2} }
	
	\item All $x_i^*$, $i=1,2,...,c$ among all clusters in the network satisfy the following
	\eq{\text{\bf Global Conservation: }\quad \sum_{i=1}^{c} (A_ix_i^*-b_i)=0 \label{eq_conserv2} }
	
\end{itemize}
Let $\bm x^*=\col \{x_1^*,x_2^*,...,x_c^*\}$ be the column collection of the consensus value to which all agents in the same cluster converge.  From (\ref{eq_partition21}) and (\ref{eq_conserv2}) one has $A\bm x^*=b$. Thus the $x^*_{ij}$ satisfying the local consensus (\ref{Consensus2}) and the global conservation (\ref{eq_conserv2}) are said to form
a\emph{ conservation-consensus solution} $\bm x^*$ to $Ax=b$.

Note that here all $x^*_{ij}$ in the same cluster are the same, which is part of the solution to the overall equation, while, in contrast, in the consensus-conservation solution defined in the previous section, the $x_{ij}^*$ in each cluster $i$ jointly form a solution to the overall equation $Ax=b$.

\subsection{The Update}
In order to achieve the global conservation, we introduce an additional state $z_{ij}(t)\in \mathbb{R}^{m_{ij}}$ at each agent $i_j$. Let $E_{ij}\in \mathbb{R}^{m_{ij}\times m}$ consist of rows of the identity matrix $I_m$ such that $\col \{E_{ij},j=1,2,...,c_i\}=I_m$ and $$A_{ij}=E_{ij}A_i.$$ Let $\bm{z}_i(t)\in \mathbb{R}^{m}$ denote the column of all coordination states in cluster $i$, $i=1,2,...,c_i$, that is, $$\bm{z}_i=\col \{z_{i1},z_{i2},...,z_{ic_i}\}. $$ Then $$z_{ij}=E_{ij}\bm{z}_i.$$

Suppose each cluster $i$ is able to access its neighbor cluster's coordination state $\bm{z}_k(t)$, $k\in \mathcal{N}_i$ through cluster-cluster communication and then distributes $E_{ij}\bm{z}_k(t)$ to each agent $i_j$ within cluster $i$. Within each cluster $i$, each agent $i_j$ is able to access to its neighbors' state $x_{ik}, i_k\in \mathcal{N}_{ij}$, through agent-agent communication in cluster $i$. Then one proposes the following update for each agent $i_j$, $i=1,2,...,c$ and $j=1,2,...,c_i$, at time $t$:
\begin{align}	\label{AEFlowx}
\dot{x}_{ij}=&-A'_{ij} \!\left(\!A_{ij}x_{ij}-b_{ij}-\!\!\sum\limits_{k \in {\mathcal{N}_{i}}} \!\!\left (z_{ij}\!-\!E_{ij}\bm{z}_{k}\right)\right)\nonumber\\
&-\sum_{i_k\in \mathcal{N}_{ij}}\left({x}_{ij}-{x}_{ik}\right)\\
\dot{z}_{ij}=&~A_{ij}x_{ij}-b_{ij}-\sum\limits_{k \in {\mathcal{N}_{i}}} \left (z_{ij}-E_{ij}\bm{z}_{k}\right) \label{AEFlowz}
\end{align}
where the first line of update \eqref{AEFlowx} and \eqref{AEFlowz} aim to achieve global conservation in (\ref{eq_conserv2}) while the second line of \eqref{AEFlowx} aims to achieve the local consensus in (\ref{Consensus2}).

Note immediately that the proposed updates \eqref{AEFlowx}-\eqref{AEFlowz} are \textbf{distributed} in the sense that implementation of them only require communication of coordination states $\bm{z}_k(t)$ among cluster-neighbors and communication of states $x_{ij}(t)$ among neighbor agents within the same cluster. Moreover, compared with existing consensus-based distributed linear equation solvers \cite{SJA15TAC,SZLDA16SCL,BSUA16NACO,XSD17TIE,GBU17TAC}, distributed updates \eqref{AEFlowx}-\eqref{AEFlowz} proposed in this paper
\begin{itemize}
	\item require \textbf{much less knowledge} of the overall equation and control states of \textbf{much smaller dimension}. For a given overall linear equation with $A\in \mathbb{R}^{m\times n}$, each agent $i_j$ knows $A_{ij}\in \mathbb{R}^{m_{ij}\times n_{i}}$ and $b_{ij}\in \mathbb{R}^{m_{ij}}$, and controls states $x_{ij}(t)\in \mathbb{R}^{n_{i}}$, and $z_{ij}(t)\in \mathbb{R}^{m_{ij}}$. Sizes of these locally available matrices and state vectors could change with respect to the number of clusters and the number of agents in each cluster. From \eqref{n_im_i2} and partitions in Fig. \ref{S2Demo}, one has that increasing $c_i$ and $c$ leads to the decreases of $m_{ij}$ and $n_{i}$, respectively. Specially, when the number of clusters is $n$ and the number of agents within each cluster is $m$, that is, $c=n$ and $c_i=m$, each agent only needs to know \textbf{two scalar entries} $A_{ij}\in \mathbb{R}$, $b_{ij}\in \mathbb{R}$ and  updates \textbf{two scalar states}, namely $x_{ij}(t)\in \mathbb{R}, \ z_{ij}(t)\in \mathbb{R}$.
	\item allow all agents' state vectors to be of \textbf{different dimensions}, which is the same as the distributed updates \eqref{LEFlowx}-\eqref{LEFlowz} in previous section.
\end{itemize}

\subsection{Main result}
Before proceeding, we first derive a compact form of \eqref{AEFlowx}-\eqref{AEFlowz}. Towards this end, we let $\bm{x}_i\in \mathbb{R}^{c_in_i}$ denote the column collection of all agents' states in cluster $i$, $i=1,2,...,c$, that is, $$\bm x_i=\col \{x_{i1},x_{i2},...,x_{ic_i}\}. $$ Let
\begin{align}\label{defbarA}
\bar{A}_{i}=\diag\{{A}_{i1}, \cdots,  {A}_{i{c_i}}\},\quad \bar{L}_{\mathbb{G}_i}={L}_{\mathbb{G}_i}\otimes I_{n_i}
\end{align} with  ${L}_{\mathbb{G}_i}$ the Laplacian matrix of the $c_i$-node connected graph $\mathbb{G}_i$.
From equations \eqref{AEFlowx}-\eqref{AEFlowz} and $\col \{E_{ij},j=1,2,...,c_i\}=I_m$,  one has
\begin{align}	\label{S2CompactFlowx}
\dot{\bm{x}}_{i}=&-\bar{A}_{i}' \left(\bar{A}_{i}\bm{x}_{i}-{b}_{i}- \sum_{k\in \mathcal{N}_i}\left(\bm z_i- \bm z_k\right)   \right)-\bar{L}_{\mathbb{G}_i}\bm{x}_i\\
\dot{\bm{z}}_{i}=&\bar{A}_{i}\bm{x}_{i}-{b}_{i}- \sum_{k\in \mathcal{N}_i}\left(\bm z_i- \bm z_k\right)  \label{S2CompactFlowz}
\end{align} for $i=1,2,...,c$.
Let $\bm{x}=\col\{\bm{x}_1, \cdots, \bm{x}_c\}$ and $\bm{z}=\col\{\bm{z}_1, \cdots, \bm{z}_c\}$, \begin{align}
\hat{A}=\diag\{\bar{A}_{i}, \cdots,  \bar{A}_{c}\}, \quad \hat{b}=\col\{{b}_{i}, \cdots,  {b}_{c}\},\label{eq_Ahat3}\\
\hat{L}=\diag\{\bar{L}_{\mathbb{G}_1},\cdots,\bar{L}_{\mathbb{G}_c}\}, \quad \hat{L}_{\mathbb{G}}= {L}_{\mathbb{G}}\otimes I_{m}\label{eq_Ahat4}
\end{align}with
${L}_{\mathbb{G}}$ the Laplacian matrix of the $c$-node connected graph $\mathbb{G}$. Equations (\ref{S2CompactFlowx})-(\ref{S2CompactFlowz}) can be written in the following compact form:
\begin{align}	\label{S2WCompactFlowx}
\dot{\bm{x}}=&-\hat{A}' \left(\hat{A}\bm{x}-\hat{b}- \hat{L}_{\mathbb{G}}\bm{z}\right)-\hat{L}\bm{x}\\
\dot{\bm{z}}=&~\hat{A}\bm{x}-\hat{b}- \hat{L}_{\mathbb{G}}\bm{z} \label{S2WCompactFlowz}
\end{align}
which is
\begin{align}	\label{S2MCompact}
\begin{bmatrix}
\dot{\bm{x}}\\
\dot{\bm{z}}
\end{bmatrix}=Q\begin{bmatrix}
{\bm{x}}\\
{\bm{z}}
\end{bmatrix}+\begin{bmatrix}
\hat{A}'\hat{b}\\
-\hat{b}
\end{bmatrix}
\end{align}
with
\begin{align}
Q=\begin{bmatrix}\label{eq_Q2}
-\hat{A}'\hat{A}-\hat{L} & \hat{A}' \hat{L}_{\mathbb{G}}\\ \\
\hat{A} & -\hat{L}_{\mathbb{G}}
\end{bmatrix}.
\end{align}
\smallskip

\begin{theorem}\label{T2}
	Suppose $Ax=b$ has at least one solution, and the graphs $\mathbb{G}_i$, $i=1,2,...,c$ and $\mathbb{G}$ are connected and bidirectional. Then under the distributed updates \eqref{AEFlowx} and \eqref{AEFlowz}, all $x_{ij}(t)$ with $i=1,2,\cdots,c$ and $j=1,2,...,c_i$ converge exponentially fast to constant vectors $x_{ij}^*$ which satisfy the local consensus (\ref{Consensus2}) and the global conservation (\ref{eq_conserv2}) and thus form
	a conservation-consensus solution $\bm x^*$ to $Ax=b$.
\end{theorem}
\smallskip

\noindent\textbf{Proof of Theorem \ref{T2}:}  We first prove that there exists a constant vector $\col\{\hat{\bm{x}},\hat{\bm{z}}\}$ which is an equilibrium of \eqref{S2MCompact}. Since there exists a constant vector $y\in \mathbb{R}^n$ such that $Ay=b$, using the definition of $A_{i}, b_{i}$ in (\ref{eq_partition21}),
one has
$$\begin{bmatrix}
{A}_{1} & {A}_{2} & \cdots & A_{c}
\end{bmatrix}y=\sum_{i=1}^{c}{b}_{i}$$
Partition $y=\col\{y_1,\cdots,y_c\}$ with $y_i\in \mathbb{R}^{n_i}$. Then one has
$$\sum_{i=1}^{c} \left(A_iy_i-b_{i}\right)=0$$
It follows from this and (\ref{eq_partition22}) that
$$\sum_{i=1}^{c} \left(\begin{bmatrix}
{A}_{i1}\\\vdots\\{A}_{ic_i}
\end{bmatrix}y_i-\begin{bmatrix}
{b}_{i1}\\\vdots\\{b}_{ic_i}
\end{bmatrix}\right)=0$$
This and the definitions of $\bar{A}_i$ in \eqref{defbarA} imply
\begin{align}\label{compAb}
\sum_{i=1}^{c} (\bar{A}_i{\bm{y}}_i-b_i)=0
\end{align} with $\bm{y}_i=\bm{1}_{c_i}\otimes y_i$.
Let $\hat{\bm{x}}=\col\{\bm{y}_1,\cdots,\bm{y}_c\}$. From \eqref{compAb} and the definitions of $\hat{A},\hat{b}$ in (\ref{eq_Ahat3}), one has
\begin{align} \label{xspace2}
\left({\bf 1}'_{c}\otimes I_{m}\right)\left(\hat{A}\hat{\bm{x}}-\hat{b}\right)=0
\end{align}
Recall that $\hat{L}_{\mathbb{G}}=L_{\mathbb{G}}\otimes I_m$ with $L_{\mathbb{G}}$ the Lapalacian matrix of a $c$-node connected, bidirectional graph $\mathbb{G}$. Then
\begin{align}
\image \hat{L}_{\mathbb{G}}=\ker \left({\bf 1}'_{c}\otimes I_{m}\right)
\end{align}
which with \eqref{xspace2} implies
\begin{align}\label{Sxespace}
\left(\hat{A}\hat{\bm{x}}-\hat{b}\right)\in \image \hat{L}_{\mathbb{G}}.
\end{align}
Then there exists a constant vector $\hat{\bm z}$ such that
\begin{align} \label{Ax-b-lz=0}
\hat{A}\hat{\bm{x}}-\hat{b}- \hat{L}_{\mathbb{G}}\hat{\bm{z}}=0
\end{align}
In addition, since $\bar{L}_{\mathbb{G}_i}={L}_{\mathbb{G}_i}\otimes I_{n_i}$  with ${L}_{\mathbb{G}_i}\otimes I_{n_i}$ the Laplacian matrix of the $c_i$-node connected graph $\mathbb{G}_i$, and since $\bm{y}_i=\bm{1}_{c_i}\otimes y_i$, one has
$$\bar{L}_{\mathbb{G}_i}{\bm{y}}_i=0,\quad i=1,2,...,c.$$
Then because $\hat{L}=\diag\{\bar{L}_{\mathbb{G}_1},\cdots,\bar{L}_{\mathbb{G}_c}\}$ and $\hat{\bm{x}}=\col\{\bm{y}_1,\cdots,\bm{y}_c\}$, one has
\begin{align} \label{Lx0}
\hat{L}\hat{\bm{x}}=0
\end{align}
together with (\ref{Ax-b-lz=0}), this implies that $\col\{\hat{\bm x},\hat{\bm z}\}$ is an equilibrium of \eqref{S2MCompact}.
\smallskip

Second, we analyze the convergence of the error
\begin{align}
\bm{e}(t)=\begin{bmatrix}
\bm{x}(t)\\
\bm{z}(t)
\end{bmatrix}-\begin{bmatrix}
\hat{\bm x}\\
\hat{\bm z}
\end{bmatrix}.
\end{align}
From \eqref{S2MCompact} and the fact that $\col\{\hat{\bm x},\hat{\bm z}\}$ is an equilibrium of \eqref{S2MCompact}, one has
\begin{align}\label{eupdate2}
\bm{\dot e}=Q\bm{e}
\end{align}
From Lemma \ref{Lemma_eig}, the structure of $Q$ in (\ref{eq_Q2}) and the fact that Laplacian matrices $\hat{L}$ and $\hat{L}_{\mathbb{G}}$ are symmetric and positive semi-definite, one has all eigenvalues of $Q$ are real negative or 0. Moreover, if $0$ is an eigenvalue of $Q$, it must be non-defective. Thus there exists a constant vector $q\in \ker Q$ such that $\bm{e}(t)$ of the linear time-invariant error system (\ref{eupdate2}) converges to $q$ exponentially fast\cite{HAMLinearSys}. Thus  $\col \{\hat{\bm x}(t),\hat{\bm z}(t)\}$ converges exponentially fast to a constant vector $\col\{\hat{\bm x}^*, \hat{\bm z}^*\}$, where
\eq{\label{eq_xzs2}\begin{bmatrix}
		\hat{\bm x}^*\\
		\hat{\bm z}^*
	\end{bmatrix}=\begin{bmatrix}
	\hat{\bm x}\\
	\hat{\bm z}
\end{bmatrix}
+q,\quad q\in \ker Q.} Partition the constant vector $\hat{\bm x}^*$ such that \eq{\label{eq_xx}\hat{\bm{x}}^*=\col\{\bar{{\bm x}}^*_1, \cdots, \bar{{\bm x}}^*_{c}\}} where $\bar{\bm x}_i^*\in \mathbb{R}^{c_in_i}$ is further partitioned as \eq{\bar{\bm{x}}_i^*=\col \{x_{i1}^*,x_{i2}^*,...,x_{ic_i}^*\} \label{eq_xxx2}} with $x_{ij}^*\in \mathbb{R}^{n_i}$. Evidently, $x_{ij}(t)$ converges to $x_{ij}^*$ exponentially fast. In the following one only needs to show that all these $x_{ij}^*$ satisfy the local consensus (\ref{Consensus2}) and the global conservation (\ref{eq_conserv2}).

From (\ref{eq_xzs2}) and the property that $\col\{\hat{\bm x},\hat{\bm z}\}$ is an equilibrium of \eqref{S2MCompact}, one has $\col\{\hat{\bm x}^*,\hat{\bm z}^*\}$ is an equilibrium of \eqref{S2MCompact}. Then
\begin{align}	\label{S3WCompactFlowx}
0=&-\hat{A}' \left(\hat{A}\hat{\bm{x}}^*-\hat{b}- \hat{L}_{\mathbb{G}}\hat{\bm{z}}^*\right)-\hat{L}\hat{\bm{x}}^*\\
0=&~\hat{A}\hat{\bm{x}}^*-\hat{b}- \hat{L}_{\mathbb{G}}\hat{\bm{z}}^* \label{S3WCompactFlowz}
\end{align} It follows that \eq{\label{eq_Lap}\hat{L}\hat{\bm x}^*=0}
From this, (\ref{eq_xx}) and the definition of $\hat{L}$, one has \eq{\bar{L}_{\mathbb{G}_i}\bar{\bm x}_i^*=0, \quad i=1,2,...,c.} Note that $\bar{L}_{\mathbb{G}_i}=L_{\mathbb{G}_i}\otimes I_{n_i}$ with $L_{\mathbb{G}_i}$ the Laplacian matrix of a connected bidirectional graph $\mathbb{G}_i$. Then there must be a constant vectors $x_i^*\in \mathbb{R}^{n_i}$ such that \eq{\label{Consensus3}\bar{\bm x}_i^*=\bm{1}_{c_i}\otimes x^*_i.} This and (\ref{eq_xxx2}) imply \eq{\label{eq_consensusL}x^*_{i1}=x^*_{i2}=\cdots=x^*_{ic_i}=x^*_i} for $i=1,2,...,c$.

From (\ref{Consensus3}), the partition of $A_i$ in (\ref{eq_partition22}) and $\bar{A}_i=\diag\{{A}_{i1}, \cdots,  {A}_{i{c_i}}\}$ in (\ref{defbarA}), one has \eq{\label{eq_bAA}\bar{A}_i \bar{\bm x}_i^*=A_ix_i^*}
From (\ref{S3WCompactFlowz}) and the definitions of $\hat{A},\hat{b}, \hat{L}_{\mathbb{G}}$, one has \eq{\label{eq_T21}\matt{\bar{A}_1\bar{{\bm x}}^*_1\\ \bar{A}_2\bar{{\bm x}}^*_2\\ \vdots \\ \bar{A}_c\bar{{\bm x}}^*_c}-\matt{b_1\\ b_2\\ \vdots \\ b_c}-(L_\mathbb{G}\otimes I_m)\hat{z}=0}
This equality and (\ref{eq_bAA}) imply \eq{\label{eq_T22}\matt{A_1x_1^*\\ A_2x_2^*\\ \vdots \\ A_cx_c^*}-\matt{b_1\\ b_2\\ \vdots \\ b_c}-(L_\mathbb{G}\otimes I_m)\hat{z}=0}
Premultiplying by $\bm{1}_c'\otimes I_m$ on both sides of (\ref{eq_T22}), one has
\eq{\label{eq_conserv03}\sum_{i=1}^{c} (A_ix_i^*-b_i)-[({\bf 1}_c'L_\mathbb{G})\otimes I_m]\hat{z}=0.}
Note that $L_\mathbb{G}$ is the Laplacian matrix of a $c$-node connected and bidirectional graph $\mathbb{G}$, one has ${\bf 1}_c'L_\mathbb{G}=0$. Thus \eq{\label{eq_conserv3}\sum_{i=1}^{c} (A_ix_i^*-b_i)=0.}

From (\ref{eq_consensusL}) and (\ref{eq_conserv3}), one sees all $x_{ij}^*$ satisfy the local consensus (\ref{Consensus2}) and the global conservation (\ref{eq_conserv2}). Therefore all $x_{ij}(t)$ converge to constant vectors and thus form
a conservation-consensus solution $\bm x^*$ to $Ax=b$. This completes the proof. $\qed$

\subsection{Simulations}
We utilize the double-layer network as in Fig. \ref{Fig1} to solve the linear equation $A\bm{x}=b$, which is partitioned according to the structure in Fig. \ref{S2Demo} with details as follows:

\begin{figure}[h]
	\centering
	\includegraphics[width=8 cm]{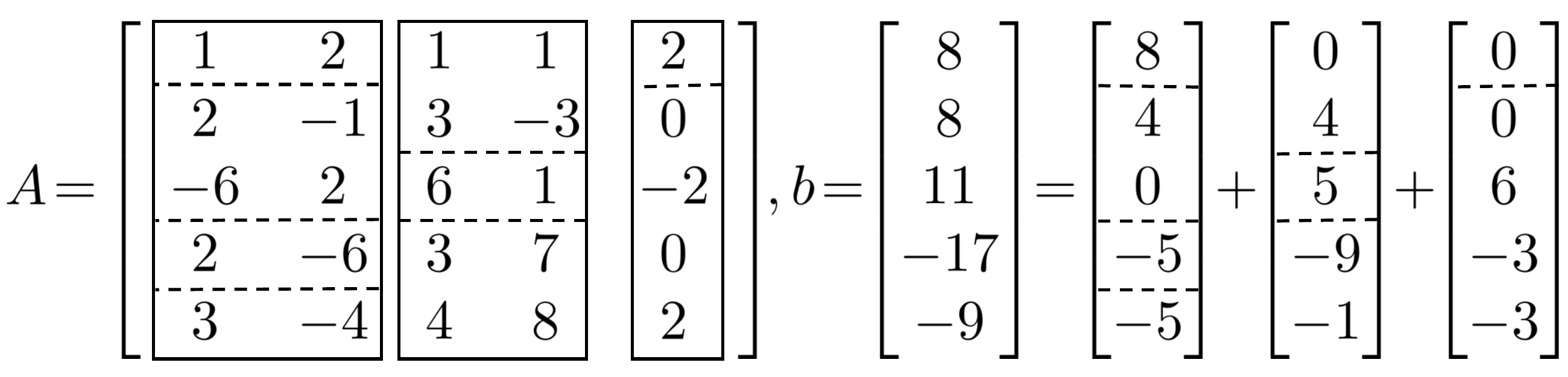}
\end{figure}

\begin{figure}[h]
	\centering
	\includegraphics[width=8.5 cm]{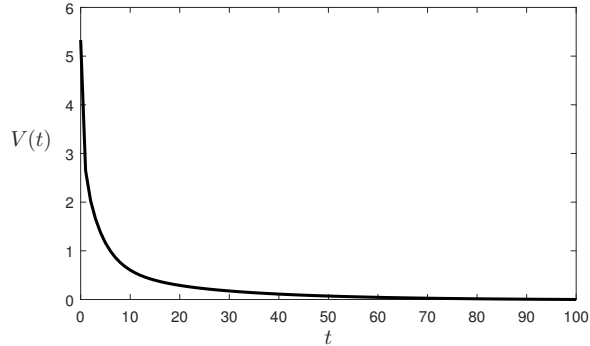}
	\caption{Evolution of $V(t)$ under distributed updates \eqref{AEFlowx}-\eqref{AEFlowz}}
	\label{Fig4}
\end{figure}

Suppose each agent $i_j$ knows $A_{ij}$ and $b_{ij}$, and employ the updates \eqref{AEFlowx} and \eqref{AEFlowz} with arbitrary initializations . Let $\bm{x}^*=\col \{x_1^*,x_2^*,x_3^*\}$ where $x_1^*=\col \{0.84,2.87\}$, $x_2^*=\col \{1.99,-1.07\}$, and $x_3^*=0.25$. Then $\bm x^*$ is a solution to $Ax=b$. Let
$$V(t)=\dfrac{1}{2}\sum_{i=1}^3\sum_{j=1}^{c_i}\|{x}_{ij}(t)-{x}_i^*\|^2_2$$ Then $V(t)$ measures the closeness of all agent states to forming a conservation-consensus solution. Simulations shown in Fig. \ref{Fig4} suggest that $V(t)$ converges exponentially fast to 0, which indicates all $x_{ij}(t)$ converge exponentially fast to constant vectors that form a consensus-conservation solution $\bm x^*$. This is consistent with Theorem \ref{T2}.

\section{Conclusion}\label{Sec_Con}
This paper has devised distributed algorithms in a double-layered multi-agent framework for solving linear equations, which consists of clusters and each cluster is composed of a network of agents. In these distributed algorithms, each agent is not required to know as much as a complete row or column of the overall linear equation. Both analytical proof and simulation results are provided to validate exponential convergence. Future work includes generalization of the proposed algorithms to time-varying directed networks, application to achieving least square solutions, investigation of the impact when different weights are assigned to the conservation and consensus, and distributed algorithms in networks of more than two layers.

\section{Appendix} \label{Sec_Appendix}
\noindent{\bf Proof of Lemma \ref{Lemma_eig}:}
Let $\lambda$ denote any eigenvalue of $M$ with a non-zero eigenvector $\col\{\bm{u}, \bar{\bm u}\}$. Then
\begin{align} \label{Eigenvector}
M\begin{bmatrix}
\bm{u}\\\bar{\bm{u}}
\end{bmatrix}=\lambda\begin{bmatrix}
\bm{u}\\\bar{\bm{u}}
\end{bmatrix}
\end{align} with $$M=\begin{bmatrix}
-M_1'M_1-M_2& M_1' M_3\\ \\
M_1 & -M_3
\end{bmatrix}.$$ Let $\bar{M}=\begin{bmatrix}
I&0\\
0&M_3'
\end{bmatrix}M$. Then one has
$$\bar{M}=\begin{bmatrix}
-M_1'M_1-M_2 & M_1' M_3\\ \\
M_3'M_1 & -M_3'M_3
\end{bmatrix}$$ which can be written as \eq{\label{eq_bM}\bar{M}=-\begin{bmatrix}
		M_1' \\ -M_3'
	\end{bmatrix}\!\!\begin{bmatrix}
	M_1 & -M_3
\end{bmatrix}-\begin{bmatrix}
M_2& 0\\0&0
\end{bmatrix}.} Thus $\bar{M}$ is negative semi-definite.
Premultiplying by $\begin{bmatrix}
\bm{u}\\\bar{\bm{u}}
\end{bmatrix}'\begin{bmatrix}
I&0\\
0&M_3'
\end{bmatrix}$ on both sides of (\ref{Eigenvector}), one has
\begin{align}
\begin{bmatrix}
\bm{u}\\\bar{\bm{u}}
\end{bmatrix}'
\bar{M}\begin{bmatrix}
\bm{u}\\\bar{\bm{u}}
\end{bmatrix}=\lambda
\begin{bmatrix}
\bm{u}\\\bar{\bm{u}}
\end{bmatrix}'
\begin{bmatrix}
I& 0\\0&M_3
\end{bmatrix}
\!
\begin{bmatrix}
\bm{u}\\\bar{\bm{u}}
\end{bmatrix} \label{ND-equ}
\end{align}

First, we prove that $\lambda$ must be real by contradiction. Suppose $\lambda =\alpha+\beta\bm{i}$ where $\beta\neq 0$. Since $\bar{M}$ is negative semi-definite, then the imaginary part of the left-hand side of (\ref{ND-equ}) is 0. So therefore is the imaginary part of the right-hand side. It follows that
$$\beta
\begin{bmatrix}
\bm{u}\\\bar{\bm{u}}
\end{bmatrix}'
\begin{bmatrix}
I& 0\\0&M_3
\end{bmatrix}
\begin{bmatrix}
\bm{u}\\\bar{\bm{u}}
\end{bmatrix}=0$$
Since $\beta\neq 0$ there follows
$$\bm u'\bm u+\bar{\bm u}'M_3\bar{\bm u}=0.$$
Recall that $M_3$ positive semi-definite. Hence
$$\bm u=0,\quad M_3\bar{\bm u}=0$$
Taken with \eqref{Eigenvector} and noting $\lambda\neq 0$ since $\beta\neq 0$, one has $\bar{\bm u}=0$. This and the assumption that  $\bm u=0$ contradict to the fact that $\col \{\bm u, \bar{\bm u}\}$ is non-zero. Thus $\beta=0$.
Therefore, $\lambda$ is real. From this, \eqref{ND-equ}, $\bar{M}$ is negative semi-definite and $M_3$ is positive semi-definite, one has $$\lambda \le0.$$

Second, if $\lambda=0$ is an eigenvalue of $M$, we prove that it must be non-defective by contradiction. Suppose $\lambda=0$ is defective, then there exists a non-zero vector $\col\{\bm{v},\bar{\bm{v}}\}$  such that
\begin{align} \label{eq_eig}
M\begin{bmatrix}
\bm{v} \\
{\bar{\bm{v}}}
\end{bmatrix} = \begin{bmatrix}\bm{u} \\
{\bar{\bm{u}}}
\end{bmatrix}
\end{align}
and
\begin{align} \label{eq_eig2}
M\begin{bmatrix}
\bm{u} \\
{\bar{\bm{u}}}
\end{bmatrix} = 0
\end{align}
Premultiplying by $\begin{bmatrix}
\bm{u}\\\bar{\bm{u}}
\end{bmatrix}'\begin{bmatrix}
I&0\\
0&M_3'
\end{bmatrix}$ on both sides of \eqref{eq_eig}, one has \begin{align}
\begin{bmatrix}\label{mtx_mv}
\bm{u}\\\bar{\bm{u}}
\end{bmatrix}'\bar{M}\begin{bmatrix}
\bm{v}\\\bar{\bm{v}}
\end{bmatrix}\!=\!\left(\bm{u}'\bm{u}+\bar{\bm{u}}'M_3'\bar{\bm{u}}\right)
\end{align}
Premultiplying by $\begin{bmatrix}
I&0\\
0&M_3'
\end{bmatrix}$ on both sides of \eqref{eq_eig2}, one has
\begin{align} \label{mtx_mu}
\bar{M}\begin{bmatrix}
\bm{u}\\\bar{\bm{u}}
\end{bmatrix}\!=0
\end{align}
This and the fact that $\bar{M}$ is symmetric imply that the left hand side of \eqref{mtx_mv} is 0. Then
\begin{align}
\left(\bm{u}'\bm{u}+\bar{\bm{u}}'M_3'\bar{\bm{u}}\right)=0
\end{align} from which, using the fact that $M_3$ is positive semi-definite, one has \eq{\label{eq_bu0}\bm{u}=0, \quad M_3'\bar{\bm{u}}=0.}
Premultiplying by $\begin{bmatrix}
\bm{v}\\\bar{\bm{v}}
\end{bmatrix}'\begin{bmatrix}
I&0\\
0&M_3'
\end{bmatrix}$ on both sides of \eqref{eq_eig} one has $$\begin{bmatrix}
\bm{v}\\\bar{\bm{v}}
\end{bmatrix}'\bar{M}\begin{bmatrix}
\bm{v}\\\bar{\bm{v}}
\end{bmatrix}\!=\matt{\bm v'\bm u \\ \bar{\bm v}'M_3'\bar{u}}$$
The right-hand side is 0 by (\ref{eq_bu0}). Thus
\begin{align}\label{mtx_vmv} \begin{bmatrix}
\bm{v}\\\bar{\bm{v}}
\end{bmatrix}'\bar{M}\begin{bmatrix}
\bm{v}\\\bar{\bm{v}}
\end{bmatrix}=0,
\end{align}
Together with (76), this yields
\eq{\label{eq_bu01}M_1\bm{v} -M_3\bar{\bm{v}}=0, \quad M_2\bm{v}=0.} From this and the definition of $M$, one has $$M\begin{bmatrix}
\bm{v} \\
{\bar{\bm{v}}}
\end{bmatrix}=0,$$
By (\ref{eq_eig}), this yields $\col\{\bm{u},\bar{\bm{u}}\}=0$, contradicting the assumption that $\col\{\bm{u},\bar{\bm{u}}\}$ is a non-zero eigenvector. Thus, $\lambda=0$ is non-defective. \qed

\bibliographystyle{IEEEtran}
\bibliography{Shaoshuai,ConsensusGossip,CyberSecurity,FormationControl,OptiCompu,Others}

\end{document}